# Mid-Infrared Photothermal - Fluorescence in Situ Hybridization for Functional Analysis and Genetic Identification of Single Cells

Yeran Bai[†ac], Zhongyue Guo[†bc], Fátima C. Pereira[d], Michael Wagner[*de] and Ji-Xin Cheng[*abc]

[a]Department of Biomedical Engineering, Boston University, Boston, MA 02215, USA
[b]Department of Electrical and Computer Engineering, Boston University, Boston, MA 02215, USA
[c]Photonics Center, Boston University, Boston, MA 02215, USA
[d]Centre for Microbiology and Environmental Systems Science, Department of Microbiology and Ecosystem Science, University of Vienna, 1030 Vienna, Austria
[e]Department of Chemistry and Bioscience, Aalborg University, 9220 Aalborg, Denmark
[†]These authors contributed equally to the work
[*]Corresponding authors email: michael.wagner@univie.ac.at (M.W.); jxcheng@bu.edu (J.X.C.)

## Abstract

Simultaneous identification and metabolic analysis of microbes with single-cell resolution and high throughput is necessary to answer the question of 'who eats what, when, and where' in complex microbial communities. Here, we present a mid-infrared photothermal - fluorescence in situ hybridization (MIP-FISH) platform that enables direct bridging of genotype and phenotype. Through multiple improvements of MIP imaging, the sensitive detection of isotopically-labelled compounds incorporated into proteins of individual bacterial cells became possible, while simultaneous detection of FISH labelling with rRNA-targeted probes enabled the identification of the analyzed cells. In proof-of-concept experiments, we showed that the clear spectral red shift in the protein amide I region due to incorporation of $^{13}C$ atoms originating from $^{13}C$-labelled-glucose can be exploited by MIP-FISH to discriminate and identify $^{13}C$-labelled bacterial cells within a complex human gut microbiome sample. The presented methods open new opportunities for single-cell structure-function analyses for microbiology.

## Introduction

In eukaryotic cell biology, measuring single-cell behaviour and cell-to-cell heterogeneity in a complex environment is key to understanding cellular interactions in different physiological conditions.[1-7] For microorganisms, the heterogeneity in genotypic and phenotypic traits has a direct impact on human health and the functioning of environmental microbiomes.[8-11] Consequently, the rapidly developing single-cell technologies have revolutionized microbiology.[12-16] Among omics-based analyses, single-cell metabolomics provides the most immediate and dynamic picture of the functionality of a cell but it is arguably the most difficult to measure.[17,18] Due to the small amount of metabolites present in single cells and the inability for amplification, detection sensitivity challenges are posed on metabolomics technology, especially when analysing the comparably small bacterial and archaeal cells. Additionally, as the function of a cell in a given set of physiochemical conditions is a variable and dynamic property that cannot be reliably predicted from either metabolic reconstructions or genomics data alone,[12] genotyping integrated with metabolic analysis provides a better way to understand how microorganisms interact with their biotic and abiotic environment. Therefore, technologies that help bridging genotype and phenotype of microbes at the single-cell level are in high demand.[19-22]

Vibrational spectroscopy with stable isotope probing has recently emerged as a novel platform for single-cell metabolism profiling.[23-30] Compared to mass spectrometry, vibrational spectroscopy is non-destructive and promises the compatibility with genotypic analysis.[17] For stable isotope probing, cells are either incubated with specific substrates carrying isotopically-labelled atoms (most commonly $^{13}C$, $^{15}N$, $^{18}O$, and $^{2}H$) or with compounds such as heavy water ($^{2}H_2O/D_2O$) that are incorporated by all metabolic active cells and thus serve as general activity markers[28]. The newly-anabolized biomolecules including lipids, proteins, and nucleic acids that contain the substrate-derived isotopes can be detected with single cell resolution by investigating the red-shifted vibrational peaks due to the isotopic effect. Raman spectroscopy has been successfully applied to study bacterial metabolic activities by tracking incorporation of $^{2}H$ (deuterium) from $D_2O$ or $^{13}C$ from $^{13}C$-labelled substrates into single bacterial cell biomass.[21,31] In these studies the isotope-labelled cells were simultaneously identified using fluorescence in-situ hybridization (FISH) with rRNA-targeted oligonucleotide probes. However, a spontaneous Raman spectrum from a single bacterium takes about 20 seconds to acquire, resulting in limited throughput that prevents large-scale analysis. Additionally, Raman spectroscopy is sometimes challenging to integrate with fluorescence-based genotyping methods because Raman scattering and fluorescence emission can results in spectral overlap, which then complicates spectral interpretation.[31] Recently, we reported on the combination of stimulated Raman scattering and FISH (SRS-FISH) that greatly boosted the Raman spectral acquisition speed and enabled an increase in throughput of analysed microbial cells by 2-3 orders of magnitude.[20] In this study, activities of selected human gut microbiome members after incubation with different mucosal sugars in the presence of heavy water were investigated at high-throughput. However, the direct visualization of sugar metabolism by tracking the incorporation of the $^{13}C$-labelled substrates by microbiome members has not yet been achieved with SRS-FISH. Additionally, SRS imaging required that the analysed bacteria were immersed in liquid while the two-photon fluorescence imaging used for detection of FISH-labelled cells turned out to be more efficient in dry samples to avoid photobleaching, which complicated the experimental procedures[20]. In contrast, infrared (IR) absorption can be applied to study cell metabolism[32] while not suffering from fluorescence background. It should also not require different sample conditions for optimal IR and fluorescence measurements. However, the spatial resolution of conventional IR microscopy is limited to several micrometres,[33,34] which hinders imaging of individual bacteria and co-recording of IR spectra and genotypic-informative FISH images.

The recently developed mid-infrared photothermal (MIP) imaging addresses these limitations.[35-41] In MIP, two lasers in the mid-IR and the visible region are used. When the modulated mid-IR light is absorbed by the sample, it leads to sample heating and expansion (photothermal effect). The visible beam passing through the sample redirects its propagation direction due to the photothermal effect. A far-field photosensor detects the periodically modulated probe photons and an image is created through pixel-by-pixel scanning or in a widefield manner. MIP has been successfully applied to image a range of organisms from a whole nematode *Caenorhabditis elegans* to a single virus at sub-micrometre spatial resolution.[38,39,42-47] The spectra from single bacteria have been recorded with MIP with high spectral fidelity and 290 nm spatial resolution.[48,49] Additionally, MIP signal could be detected from fluorescence intensity fluctuation for fluorophore-loaded samples.[50,51] However, so far, there's no demonstration of MIP for simultaneous bacterial FISH-genotyping and metabolic imaging via isotope probing.

Here, we present a MIP-FISH platform that enables high-throughput metabolic imaging and identification of bacteria with single-cell resolution. By using oligonucleotide probes tagged with fluorophores to target signature regions in ribosomal RNA (rRNA) genes, FISH has become an indispensable tool for rapid and direct single-cell identification of microbes.[52] In this work, we greatly improved the performance of a widefield MIP microscope through multiple optimizations such as the utilization of a nanosecond laser as probe source. We then incorporated a fluorescence module on the widefield MIP to enable a co-registered MIP and fluorescence imaging from the same cells. To demonstrate the high-throughput metabolic imaging capability of MIP-FISH, we imaged newly-synthesized protein in hundreds of *Escherichia coli* cells from $^{13}$C-labelled glucose in seconds, with single cell resolution. Simultaneous identification of bacterial taxa and metabolism profiling was demonstrated by imaging bacterial mixtures including a spiked gut microbiome sample. Collectively, our results demonstrate the capability of high-throughput microbial phenotyping of metabolism and genotyping with single-cell resolution through MIP-FISH.

## Experimental

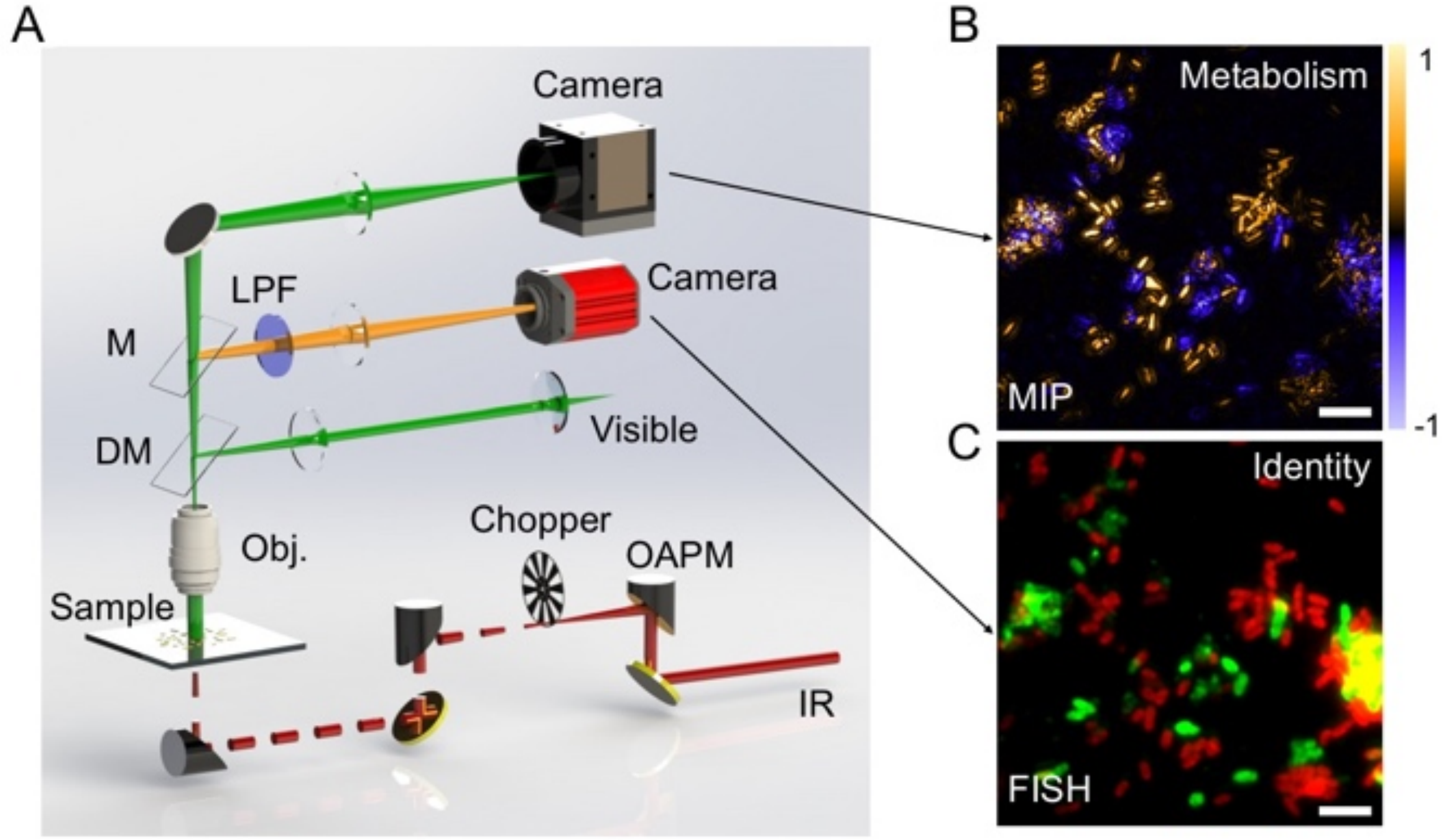

**Figure 1. Schematic of MIP-FISH for *in-situ* bacteria identification and phenotypical metabolic imaging.** (A) The setup was based on a widefield MIP with the incorporation of a fluorescence module. OAPM, off-axis parabolic mirror. DM, dichroic mirror. LPF, long pass filter. M, mirror. (B) The subtraction image of MIP signals at two IR wavenumbers provides information on cellular metabolism. Positive values (in yellow) indicate active incorporation of $^{13}$C from labelled substrates (here $^{13}$C-glucose) into protein while negative values (in blue) indicate no $^{13}$C incorporation. (C) The fluorescence image detects signal from FISH with rRNA-targeted probes, enabling the identification of bacterial taxa (*Escherichia coli* in red and *Bacteroides thetaiotaomicron* in green). Scale bars: 5 µm.

### MIP-FISH microscope

**Figure 1A** shows a schematic illustration of the MIP-FISH microscope. For MIP imaging, the mid-IR pump source is an optical parametric oscillator (OPO) laser (Firefly-LW, M Squared Lasers) with 20 ns pulse duration and 20 kHz repetition rate. The visible probe is a nanosecond laser (NPL52C, Thorlabs) with a center wavelength of 520 nm and the pulse width of 129 ns. The mid-IR beam was modulated using an optical chopper (MC2000B, Thorlabs) with a duty cycle of 50%. The mid-IR was optically chopped into pulse trains with the modulation frequency of 635 Hz and around 16 bursts of IR pulses are in the period of one camera exposure time. The microscopic objective (MPLFLN Olympus, 100X, NA 0.9) was used to focus the visible light onto the sample as well as to collect the reflected light. To record the sample scattered light for MIP imaging, a high full-well-capacity camera (Q-2HFW, Adimec) was used. For FISH imaging, a fluorescence module composed of a fluorescence camera (CS235MU, Thorlabs), dichroic beam splitter and filter sets were installed on the MIP microscope. An additional continuous wave 638 nm laser (0638-06-01-0180-100, Cobalt) was aligned with the 520 nm laser for additional fluorophore excitation. For MIP imaging, the IR power before microscope was 32.9 mW and 34.8 mW at 1612 $cm^{-1}$ and 1656 $cm^{-1}$. All presented images were normalized with IR powers at corresponding wavelengths. The visible power was less than 1 mW before microscope. Unless otherwise noted, the MIP images at one IR wavenumber were acquired at the speed of 2.4 seconds per image. For fluorescence imaging, the exposure time of the fluorescence camera was 1 second with a gain of 20dB.

### *E. coli* $^{13}$C-glucose isotope labelling and sample preparation

For data presented in Figure 2 and Figure 3, *E. coli* BW25113 was inoculated from a single colony and pre-cultured in nutrient-rich medium (either Tryptic Soy Broth or Mueller Hinton Broth) for 3 hours to reach log phase. The optical density at 600 nm was measured to estimate the concentration of cells per ml. Then cultures were diluted to a concentration around $5\times10^5$ CFU/mL in M9 minimal medium. The M9 minimal medium was supplemented with $^{12}$C-glucose or $^{13}$C-glucose, or a varying volume mixture of both, at a final concentration of 0.2% (w/v). The $^{13}$C-glucose used (D-glucose U-$^{13}C_6$, 99%, Cambridge Isotope Laboratories) was universally labelled – so that all carbon atoms were replaced with $^{13}$C atoms. Cells were harvested by centrifugation at 11,000 RPM and 4$^{o}$C for 3 minutes after 24 h of aerobic incubation at 37°C with glucose. The bacterial cells were then fixed with 10% formalin at 4$^{o}$C overnight. Multiple rounds of centrifugation and washes with deionized water were performed to remove remaining fixative. A 2 µl drop of concentrated cell solution in water was deposited on a poly-L-lysine coated IR-transparent silicon coverslip (silicon 2018, University Wafers) and dried in air.

### Multi-species sample and fluorescence in situ hybridization

For data presented in Figure 4-6, *E. coli* K-12 (DSM 498) was grown aerobically at 37°C in M9 minimal medium containing 0.4% (w/v) of either $^{12}$C-glucose (unlabelled D-glucose, 99.5%, Sigma-Aldrich) or $^{13}$C-glucose (D-glucose-$^{13}C_6$, 99%, Sigma Aldrich). Cells were grown overnight in M9 medium containing unlabelled glucose and diluted 1:100 in 5 mL of fresh medium containing either $^{12}$C or $^{13}$C -glucose. *Bacteroides thetaiotaomicron* (DSM 2079) (*B. theta*) cells were grown anaerobically (in a Coy Labs anaerobic chamber containing an atmosphere of 85% $N_2$, 10% $CO_2$ and 5% $H_2$) in Bacteroides defined minimal medium (BMM) containing 0.5% (w/v) of either $^{12}$C-glucose (unlabelled D-glucose, 99.5%, Sigma-Aldrich) or $^{13}$C-glucose (D-glucose-$^{13}C_6$, 99%, Sigma Aldrich).[53] Cells were grown overnight in BMM medium containing unlabelled glucose and diluted 1:100 in 5 mL of fresh medium containing either $^{12}$C or $^{13}$C -glucose. *E. coli* and *B. theta* cells were harvested by centrifugation at late exponential phase (9 h of growth for *E. coli* and 12 h of growth for *B. theta*), and immediately fixed in 4% formaldehyde in phosphate-buffered saline (PBS) for 2 h at 4°C. Cells were subsequently washed once with PBS, finally resuspended in 1 mL of a 50% (v/v) mixture of PBS and 96% ethanol and stored at -20°C until further use. *E. coli* cells were subsequently hybridized with the Gam42a probe tagged with Cy5 fluorophore and *B. theta* cells were hybridized with the Bac303 probe tagged with Cy3 fluorophore following a standard FISH protocol (**Supplementary Methods and Table S1**). Hybridized *E. coli* and *B. theta* cells were mixed and 2µL of this mixture was spotted on the poly-L-lysine–coated silicon coverslips and allowed to dry in air protected from light. In addition, a cultured human gut microbiome sample stored in PBS (**Supplementary Methods**) and pre-hybridized E. coli cells that have been stored in PBS:ethanol were gently mixed in an Eppendorf tube and subsequently spotted on poly-L-lysine–coated silicon coverslips. Excess of salt was removed by adding 2µL Milli-Q water onto the dried spot. Subsequently, the water was gently blown away and the spot was dried again at room temperature protected from light.

### $^{13}$C-protein replacement ratio quantification

For high-throughput single-cell analysis, the region of interests (ROIs) were determined based on the reflection images. To quantify metabolic activity, we defined the $^{13}$C-protein replacement ratio as the relative contribution of $^{13}$C-protein to the whole protein (**Figure S1**). The contribution was calculated based on four coefficients obtained from two reference samples at the original amide I (1656 cm$^{-1}$) and the shifted amide I (1612 cm$^{-1}$) bands. Since different culture and treatment conditions were used, we obtained different coefficients for in-group comparison. The details of coefficients and statistics for reported experiments are listed in **Table S2**. We also recorded the coefficients for the same reference sample on different days, and observed less than 6% variation (**Table S3**).

## Results and discussion

### A MIP-FISH platform

We improved the performance of the previously reported first-generation widefield MIP microscope[41] and achieved an over 2 orders of magnitude increase of imaging speed by making the following modifications: (1) A LED was used as the probe source in the first-generation MIP setup with a minimal pulse width of 900 ns, which was ideal for micron-sized polymer beads since the decay constant is around several microseconds.[41,54] However, the signal produced from a single bacterium is much weaker than that of a polymer bead and the decay constant is much shorter, reaching to 280 ns.[35] Therefore, we coupled a nanosecond laser with a pulse width of 126 ns to match the decay constant to improve the detection sensitivity for bacteria. (2) MIP is a shot noise limited technique and the signal to noise ratio (SNR) is proportional to the total probe photon received.[41] Therefore, we incorporated a high full-well-capacity (2 Me$^-$) camera to the current setup. (3) To accommodate the high full-well-capacity camera with the pixel size of 12 µm, we used a high magnification and high numerical aperture objective. Other improvements including shortening of the IR pathlength, galvo scanner adjusting the pointing of IR beam when tuning wavenumbers, and adding a laser speckle reduction module synergistically worked together to push the MIP imaging speed. As a comparison, we previously achieved 2 frames/s for 1 µm poly methyl methacrylate (PMMA) beads imaging with a field of view around 20 µm and a SNR of 24.[54] By implementing the optimizations, we achieved 635 frames/s for 500 nm PMMA beads imaging with a field of view of around 60 µm and a SNR of 12 (**Figure S3**). Collectively, these optimizations were essential to adapt the technique for high-throughput bacterial metabolic phenotyping.

The MIP-FISH microscope is schematically shown in **Figure 1A**. Briefly, a fluorescence module was integrated into the MIP microscope sharing the visible illumination. The mid-IR pulses are modulated with an optical chopper to a burst of pules matching the camera frame rate. In this work, we focused on imaging of the fluorophores Cy3 and Cy5, which are widely used for FISH-based detection of microbes. The 520 nm nanosecond probe source also served as the Cy3 fluorophore excitation source. The Cy5 fluorophore was excited with a second visible beam with the centre wavelength of 638 nm aligned with the 520 nm laser. Due to the different requirement of MIP and FISH imaging for cameras, we separated the two detection paths and used two cameras for recording the sample scattered light and the fluorophore-emitted light. For the bacteria taxa-specific fluorescence detection from FISH, we choose a camera with high quantum efficiency and low readout noise for the low photon-budget condition. The metabolic activity of the cells was characterized by two IR wavenumber MIP imaging (**Figure 1B**), while the identity of the cells was provided by two-colour fluorescence imaging (**Figure 1C**).

### High-throughput high-sensitivity metabolic imaging of protein synthesis in bacteria by widefield MIP

The incorporation of isotopes into cellular biomass leads to shifts in IR absorption peaks to a lower wavenumber, and has been observed for various isotopes and organisms.[32,55] The effect of $^{13}$C incorporation by cells was previously demonstrated in a point-scan MIP system.[35,56,57] Here, we use *E. coli* cells to demonstrate the capability of widefield MIP to image metabolism of $^{13}$C from isotopically-labelled compounds by bacteria. Glucose is a widely used energy source of bacteria and various amino acids can be synthesized from glucose. Therefore, we selected $^{13}$C-labelled glucose as a model substrate for this study.

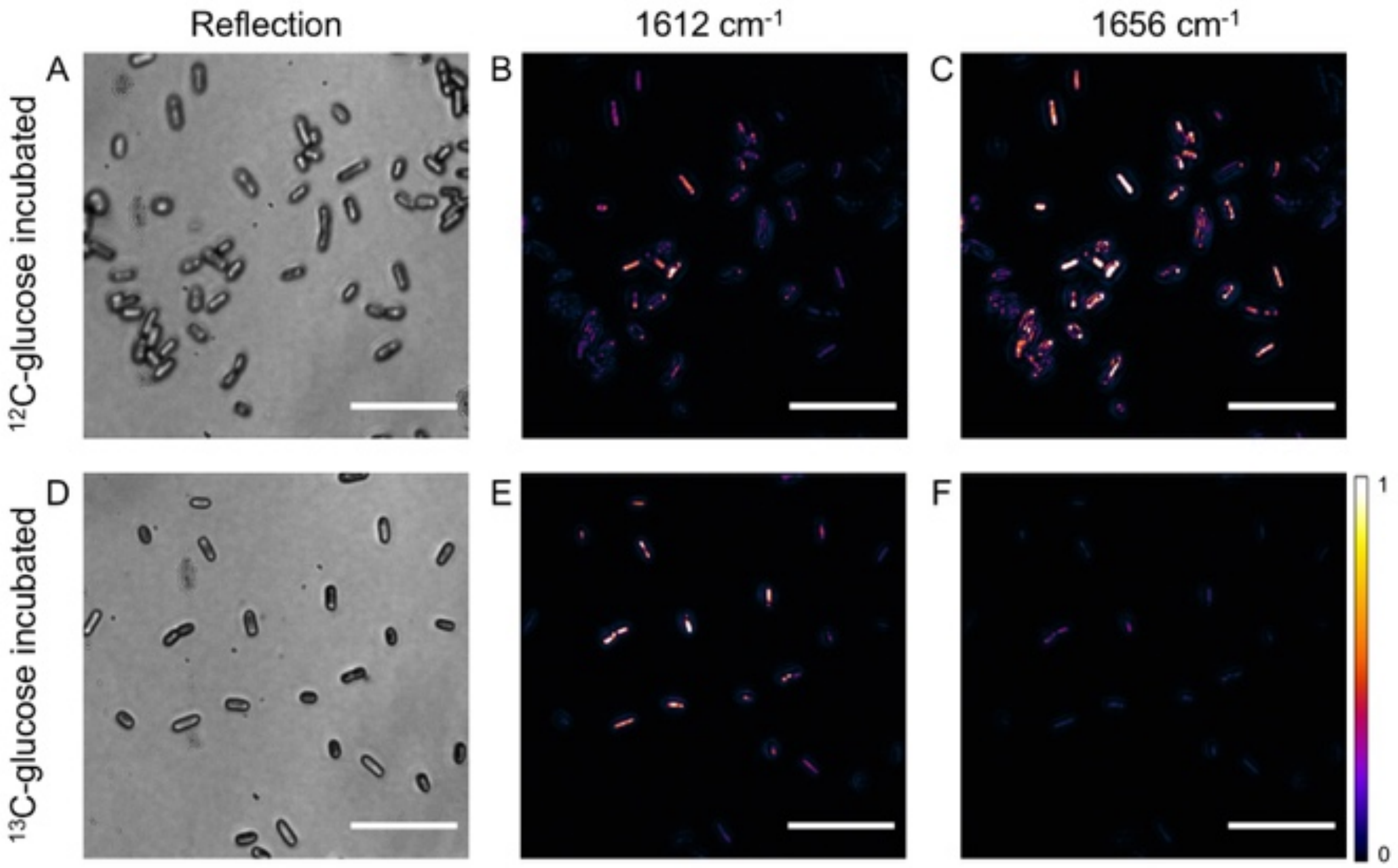

**Figure 2. Single cell metabolic imaging of $^{13}$C-glucose incorporation by widefield MIP**. Reflection and MIP images at two key protein amide I wavenumbers (1612 cm$^{-1}$ and 1656 cm$^{-1}$) for *E. coli* cells incubated with $^{12}$C-glucose (A-C) or $^{13}$C-glucose (D-F). Scale bars 10 µm.

We imaged the cells under the MIP-FISH microscope (**Figure 2**). The rod shape of the individual *E. coli* was clearly shown in both the reflection and MIP images. We acquired multispectral widefield MIP images covering the protein amide I and amide II region (1512 to 1768 cm$^{-1}$) for the unlabelled glucose ($^{12}$C-glucose) and $^{13}$C-labelled glucose ($^{13}$C-glucose) incubated cells (**Figure S2**). The $^{12}$C-glucose incubated bacteria showed a protein amide I peak at around 1656 cm$^{-1}$ while the protein amide I peak for $^{13}$C-glucose incubated bacteria was around 1612 cm$^{-1}$. The amide II peak also showed the isotopic effect with a smaller shift from 1548 cm$^{-1}$ to 1532 cm$^{-1}$. We selected two key wavenumbers representing $^{12}$C-protein (1656 cm$^{-1}$, original amide I band) and $^{13}$C-protein (1612 cm$^{-1}$, shifted amide I band), and recorded MIP images (**Figure 2 B-C, E-F**). For cell incubated with $^{12}$C-glucose, higher intensity was observed at 1656 cm$^{-1}$. For the cells incubated with $^{13}$C-glucose, higher intensity was observed at 1612 cm$^{-1}$ for shifted amide I band, indicating the incorporation of the heavier carbon atoms into the protein. Due to the high throughput of widefield MIP, we were able to acquire high-SNR MIP images of up to hundreds of bacteria within 2.4 seconds.

To quantify the percentage of $^{13}$C-protein in the whole protein pool, we defined a $^{13}$C-protein replacement ratio (**Experimental Section and Figure S1**). The estimated $^{13}$C-protein replacement ratio reaches 0.956 after 24 hours based on the residual peak at 1656 cm$^{-1}$ (**Figure S2)**, which could be considered as near full substitution.

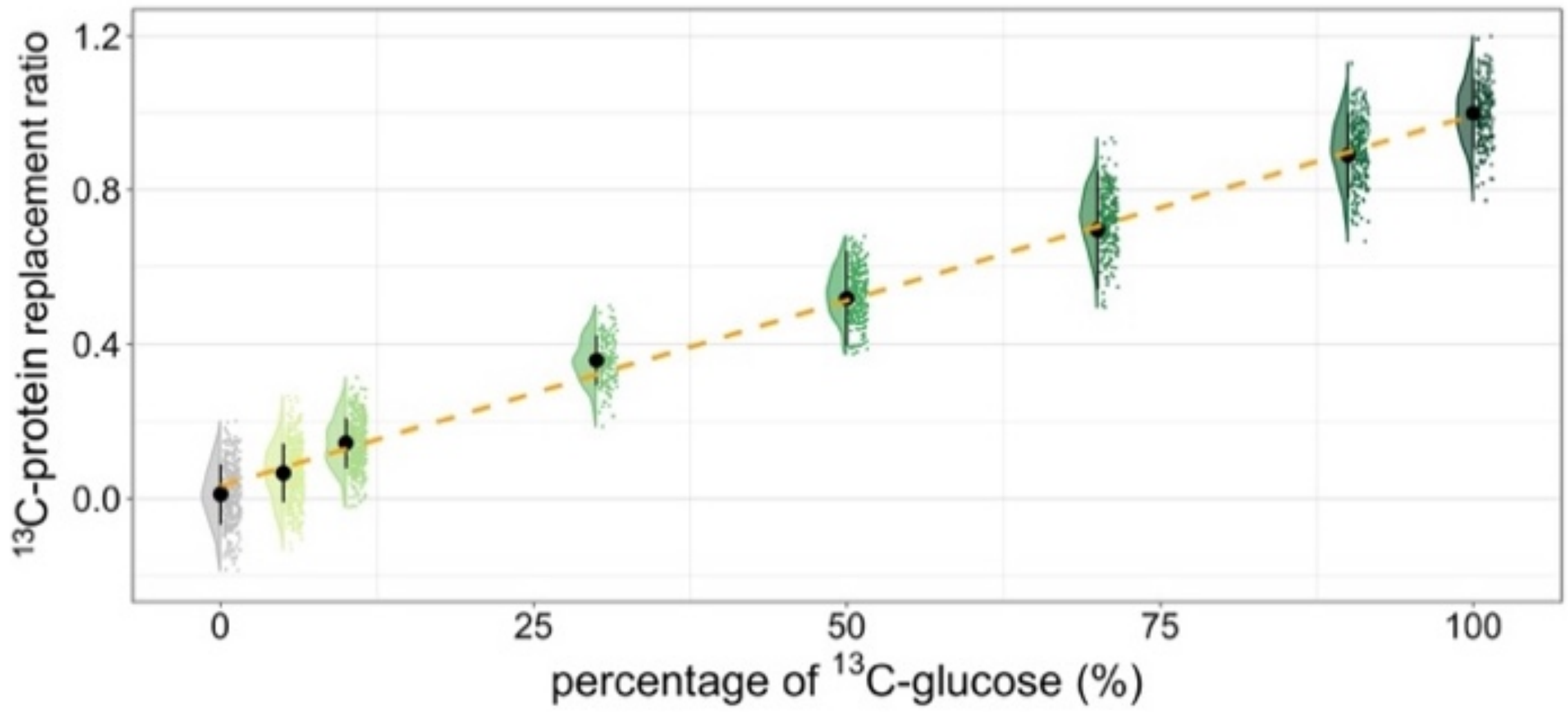

**Figure 3**. **High detection sensitivity for $^{13}$C-incorporation.** $^{13}$C-protein replacement ratio of *E. coli* cells grown for 24 hours in minimal medium supplemented with 0.2% (w/v) of glucose, at varying percentages of the total glucose in the form of $^{13}$C-glucose (0%, 5%, 10%, 30%, 50%, 70%, 90% and 100%). More than 155 cells in each group were measured to produce the mean and standard deviation. A significant difference was observed between the 0% and the 5% $^{13}$C-glucose incubated cells (pairwise t-test, p = 3.86e-22). A linear regression between percentage of $^{13}$C-glucose and $^{13}$C-protein replacement ratio is shown as a dashed line (R2 = 0.9982).

We further demonstrate the high detection limit of isotope incorporation for MIP. We incubated *E. coli* cells for 24 hours with varying percentages of $^{13}$C-glucose contributing to the total pool of available carbon. The $^{13}$C-protein replacement ratio was calculated for each group and plotted as the function of percentage of $^{13}$C-glucose in the medium (**Figure 3**). The $^{13}$C-protein replacement ratio increased as the percentage of $^{13}$C-glucose increased, and there was a clear linear correlation ($R^2$ = 0.9982) between the concentration of $^{13}$C-glucose and the incorporation of $^{13}$C into the cellular protein. Notably, a significant difference was observed for 0% and 5% $^{13}$C-glucose incubation (pairwise t-test, p = 3.86e-22). In comparison, with spontaneous Raman spectroscopy a detection limit of 8% has been described for recording deuterium incorporation in microbial biomass.[21] It should be noted that a relative low percentage of heavy water in Raman-based measurement is used to avoid potential inhibitory effects.[21] Here, we observed no differences in growth or cell morphology that could reflect toxicity, even when all carbon source available was in the form of $^{13}$C-glucose (100% $^{13}$C-glucose). This is in agreement with literature reporting that incorporation of $^{13}$C-glucose shows negligible influence on cell metabolism and physiology.[58]

Heterogeneity in $^{13}$C incorporation was observed for $^{13}$C-protein replacement ratio within each individual incubation group, despite the fact that cells were derived from an isogenic microbial population. To understand the origin of this heterogeneity, we performed multispectral MIP imaging on standard samples including polymer beads (PMMA beads 500 nm in diameter) and on a bovine serum albumin (BSA) film and calculated the mock ratio by applying a similar analysis as for $^{13}$C-protein replacement ratio (**Figure S4 and Table S4**). The standard deviation for these mock ratios from the PMMA beads and BSA film is 5 times smaller than that of the *E. coli* samples, indicating the $^{13}$C-protein replacement ratio fluctuation originated indeed mostly from phenotypic heterogeneity. This is not unexpected in batch incubations with the resulting physicochemical differences.

**Microbial identification and metabolism analysis with MIP-FISH**

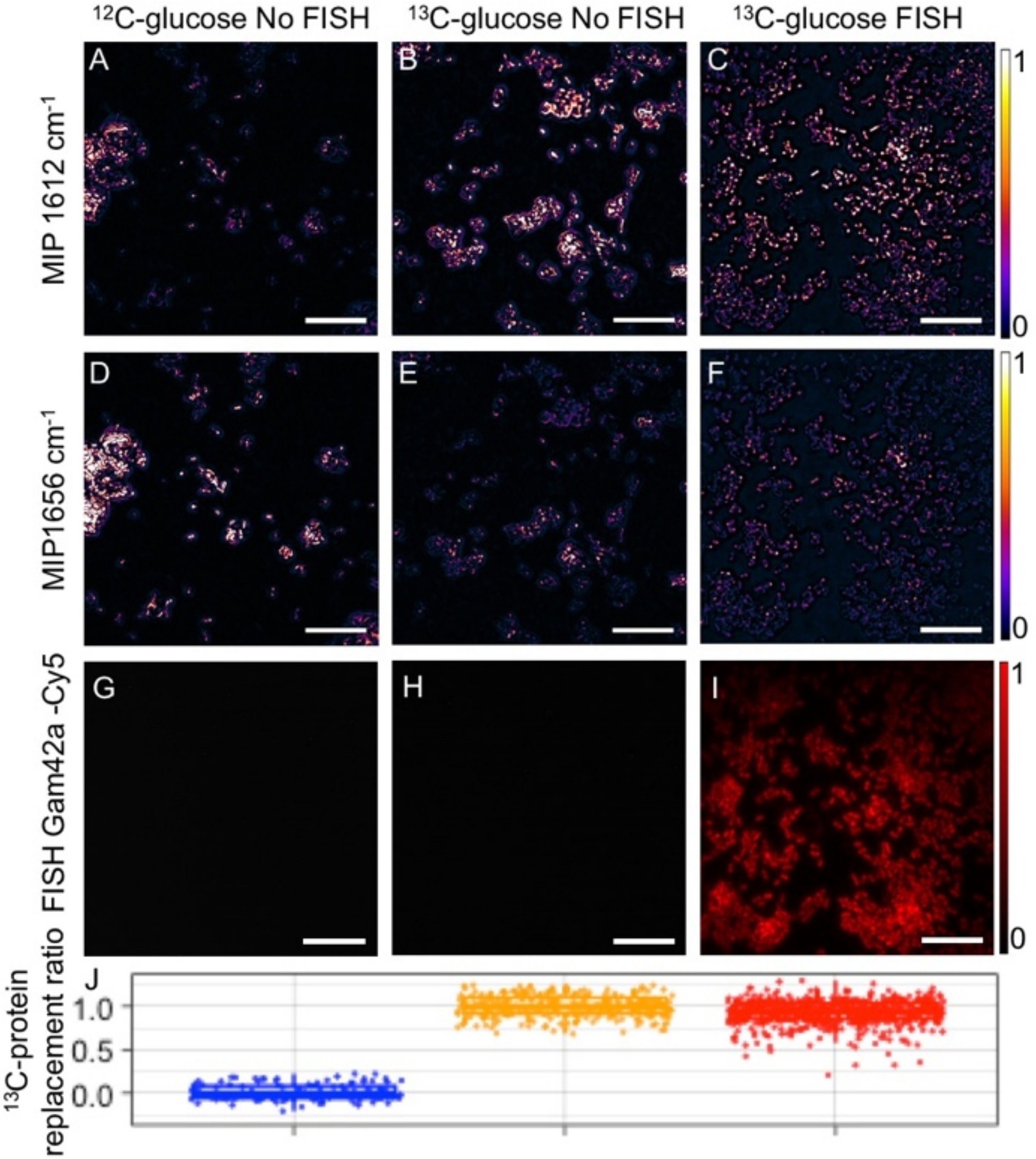

**Figure 4. FISH is compatible with MIP metabolic imaging.** *E. coli* cells grown in $^{12}$C-glucose-containing medium with no FISH labelling were imaged at the two amide I peak wavenumbers as well as by recording fluorescence in the Cy5 channel (A,D,G). $^{13}$C-glucose incubated *E. coli* with and without FISH labelling were imaged at the same channels (B,E,H and C,F,I) using identical settings. (J) The $^{13}$C-protein replacement ratio was calculated for each incubation group. A slight reduction (7.5%) was observed for the $^{13}$C-protein replacement ratio between groups with and without FISH labelling. Scale bars 10 µm.

To evaluate the capacity of MIP-FISH to simultaneously retrieve information on cellular metabolism and identity of the analyzed bacterial cells, we initially imaged *E. coli* cells that were stained by FISH with the oligonucleotide probe Gam42a-Cy5 (**Figure 4, Table S1**). Hybridized cells showed a clear signal on the fluorescence Cy5 channel that overlaps well with the MIP images. As expected, in control experiments no fluorescence signal could be detected in cells that were not hybridized. We observed for hybridized as well as non-hybridized cells higher IR intensities at 1612 cm$^{-1}$ for cells grown in $^{13}$C-glucose-containing medium. (**Figure 4**). By calculating the $^{13}$C-protein replacement ratio, a difference in cells grown with unlabelled glucose and $^{13}$C-glucose was observed, as expected. However, the FISH process slightly reduced the $^{13}$C-protein replacement ratio (7.5% on average; **Figure 4J**). We also observed a higher than 0 $^{13}$C- protein replacement ratio for $^{12}$C-glucose incubated cells with FISH labelling (**Figure S5**). One potential reason can be that the selective loss of cellular protein during the hybridization and washing steps of the FISH protocol leads to overall amide I intensity decrease, changing the quantification coefficients. An effect of the FISH protocol on quantification of isotope incorporation within microbial cells has been previously reported for other vibrational spectroscopy-based methods.[20,21] Therefore, we acquired coefficients for different cultures and treatment conditions to obtain a more accurate result (**Table S2**).

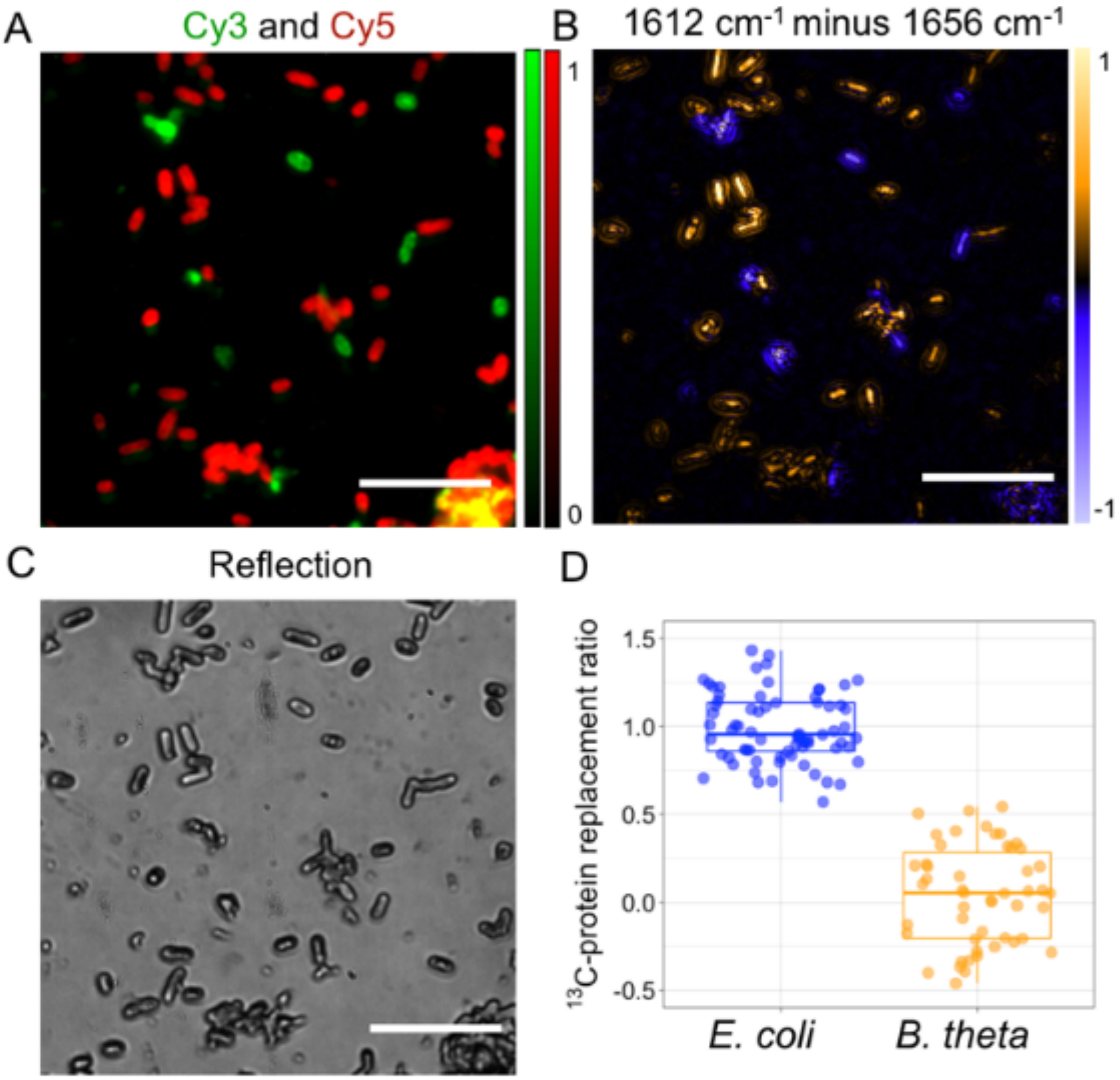

**Figure 5. MIP-FISH imaging of bacterial mixtures.** *E. coli* cells were incubated with 0.4% (w/v) $^{13}$C-glucose and hybridized with Gam42a-Cy5 oligonucleotide probe, while *B. theta* cells grown in the presence of $^{12}$C-glucose were hybridized with a Bac303-Cy3 oligonucleotide probe. (A) Fluorescence imaging for identification of *E. coli* (red) and *B. theta* (green). Scale bars 10 µm. (B) Subtraction of two MIP images (intensity at 1612 cm$^{-1}$ minus intensity at 1656 cm$^{-1}$) showed that a portion of the cells have incorporated $^{13}$C into the protein (in yellow) while other cells showed no $^{13}$C labelling (in purple). (C) Reflection image shows cell morphology of the bacterial mixture. (D) Quantification of $^{13}$C-protein replacement ratio. (Pairwise t-test: p=9.74e-33).

We further tested the capacity of MIP-FISH to identify bacterial taxa and their metabolic status on multi-species samples. We started by using an artificial mixture of two common human gut microbiome members: *E. coli* and *Bacteroides thetaiotaomicron (B. theta)*. *E. coli* cells grown in the presence of $^{13}$C-glucose were hybridized with the Gam42a-Cy5 probe, while *B. theta* cells were grown with $^{12}$C-glucose and hybridized with the Bac303-Cy3 probe (**Table S1**). Subtraction of MIP images at 1612 $cm^{-1}$ and 1656 $cm^{-1}$ revealed that a fraction of the cells on this two-species sample displayed positive subtraction values (**Figure 5B, yellow color**) indicative of $^{13}$C-glucose incorporation, while the majority of the remaining cells displayed negative values (**Figure 5B, blue color**). From the subtraction results and the growth conditions, we inferred that cells with positive contrast were *E. coli* and the cells with negative contrast were *B. theta*. However, since both *E. coli* and *B. theta* are rod-shaped bacteria of similar size, we were not able to differentiate them based on morphology (**Figure 5C**). Benefiting from the fluorescence imaging capability of MIP-FISH and the ability of rRNA-targeted FISH to discriminate bacterial taxa, we could confirm that cells with positive contrast were *E. coli*, as these displayed a Cy5 fluorescent signal resulting from hybridization with the Gam42a-Cy5 probe (**Figure 5A,** red color). In contrast, *B. theta* cells displaying Cy3 signal that originated from hybridization with the Bacteroidales probe Bac303-Cy3 (**Figure 5A,** green color) exhibited negative subtraction values. Finally, the $^{13}$C-protein replacement ratio was calculated (**Figure 5D**) and showed a significant difference between the two species (pairwise t-test, p=9.74e-33). A similar differentiation was observed for the opposite combination (*E. coli* incubated in $^{12}$C-glucose and FISH-labelled with Cy5, *B. theta* incubated in $^{13}$C-glucose and FISH-labelled with Cy3, **Figure S6**). Our results demonstrated that MIP-FISH is suitable to efficiently distinguish cells with $^{13}$C-induced protein peak shifts in mixed samples.

To test the performance of MIP-FISH beyond simple mixtures of bacteria and to demonstrate that it can be applied to identify microbes and retrieve metabolic information in a complex microbiome sample, we imaged a mixture of $^{13}$C-labelled *E. coli* and a human gut microbiome sample. The human large intestine is inhabited by trillions of gut microbes that perform a range of metabolic functions important for host health. We have therefore tested the suitability of MIP-FISH to investigate microbial metabolism of isotopically labelled compounds in such a complex setting. MIP-FISH successfully enabled us to identify *E. coli* cells grown with $^{13}$C-glucose and hybridized with a Gam42a-Cy5 probe in a gut microbiome sample that had been incubated with $^{12}$C-glucose. The 1612 $cm^{-1}$ and 1656 $cm^{-1}$ subtraction results (**Figure 6B**) and quantitative calculation of $^{13}$C-protein replacement ratio (**Figure 6D**) together with fluorescence imaging showed the assimilation of $^{13}$C-glucose into protein for *E. coli*, but not for other gut microbiome members (**Figure 6A**). As the gut microbiome contains many different microbial taxa, we performed an additional experiment in which we analyzed a much larger number of isotopically unlabelled cells from the gut microbiome (**Figure S7**). This analysis revealed a low $^{13}$C-protein replacement ratio as in **Figure 6D** showing that spectral differences between different gut microbiome members have no strong effect on the MIP-based measurements of isotopically unlabelled cells. Together, these data show the applicability of MIP-FISH to a complex microbiome sample.

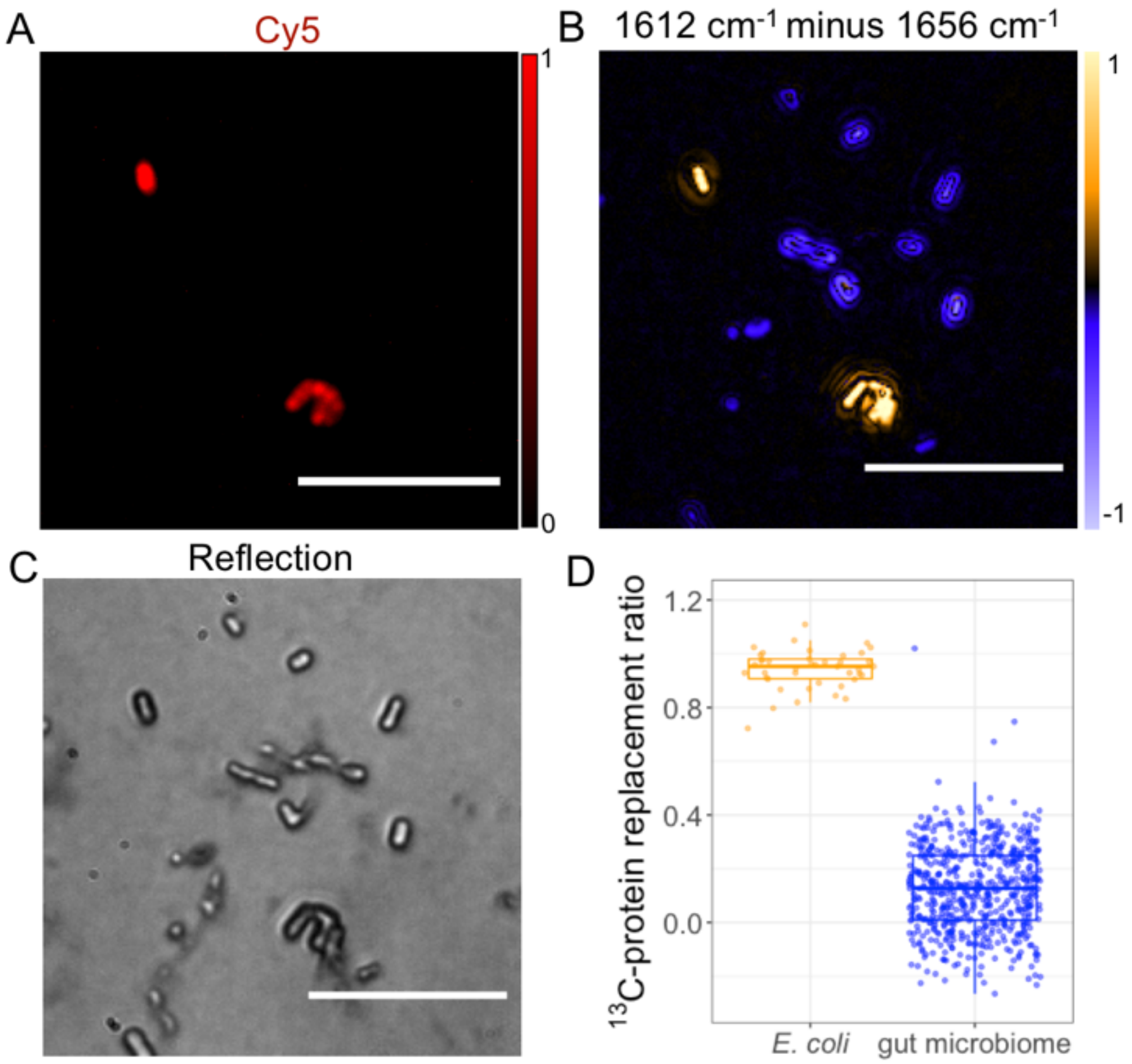

**Figure 6. MIP-FISH imaging of a gut microbiome sample with spiked $^{13}$C-labelled *E. coli* cells.** $^{13}$C-gluocse fully labelled *E. coli* cells that were FISH labelled with Cy5 were mixed with an isotopically unlabelled human gut microbiome sample. (A) Fluorescence imaging enabled localization of the added *E. coli* cells in the complex sample. Scale bars 10 µm. (B) MIP subtraction image indicated $^{13}$C-labelling for *E. coli* (in yellow). (C) Reflection image of the sample mixture. (D) Quantification of $^{13}$C-protein replacement ratio. (Pairwise t-test: p=7.69e-24).

## Conclusions

In this study, we developed a MIP-FISH platform for *in-situ* bacteria identification and metabolic imaging in a high-throughput manner with single-cell resolution. Benefiting from the high compatibility of MIP, we coupled a fluorescence module to a widefield MIP setup for fluorescence imaging of bacterial species hybridized with fluorescently-labelled oligonucleotide probes. We demonstrated the potential for applying MIP-FISH on multi-species communities and complex samples and observed good correlations between MIP metabolic imaging and FISH imaging. As a proof of concept, we successfully applied MIP-FISH to image the microbial assimilation of $^{13}$C from labelled glucose. In pure culture experiments, high sensitivity was achieved with a detection limit of 5% of $^{13}$C in total carbon. In the complex microbiome sample containing many different unlabelled microbial taxa with different chemical cellular composition, the background signal in the selected regions for the MIP-based detection of $^{13}$C-incorporation into proteins was higher, but unambiguous detection of spiked fully labelled *E. coli* cells was nevertheless possible. Microbial taxa with different physiologies and in distinct environments may display variable $^{13}$C incorporation levels and may never achieve full $^{13}$C-labelling. Under these circumstances it would be important to first evaluate if MIP-FISH is able to unambiguously discriminate $^{13}$C-labelled and $^{12}$C-labelled cells of a taxa of interest in the context of a complex community.

Keeping in mind that the protein amide II band involves nitrogen, in future studies MIP-FISH might also be suitable to study the assimilation of nitrogen-containing compounds using $^{15}$N-labelled substrates. More generally, this newly developed platform should now be ready to interrogate assimilation of many key substrates by human gut microbiome members including sweeteners, prebiotics, or even human-targeted drugs.

It is worth to compare MIP-FISH with other single-cell vibrational spectroscopy or IR-based metabolic characterization platforms for single cell isotope probing of microbes. Single-cell spontaneous Raman spectroscopy offers full-spectral coverage but strong fluorescence background could complicate the spectral analysis.[21,31,59,60] The throughput of Raman measurement is drastically improved by SRS at the cost of more expensive and complicated instrumentation, as well as limited spectral coverage.[20] Additionally, it is harder to resolve $^{13}$C and $^{15}$N assimilation than $^{2}$H as the isotopic effect of $^{13}$C and $^{15}$N is relatively small and often buried in the Raman fingerprint region.[59] On the other hand, IR offers higher signal in the fingerprint region,[61] which makes MIP a more suitable tool for high-throughput characterization of $^{13}$C and $^{15}$N assimilation in microbial cells. Additionally, widefield MIP provides similar imaging speed as SRS with the addition of compact, cost-effective and high compatibility with fluorescence merits.

We envision that the MIP-FISH platform will amend the toolbox of microbial ecologists and microbiome researchers that aim to simultaneously investigate the identity and function of individual microbial cells. Furthermore, this technique might also be useful for rapidly determining antibiotic resistance of microbial cells in complex samples[62-67] Finally, as a non-destructive single cell analytic tool, it should be feasible in the future to integrate MIP imaging with other genotyping methods beyond FISH, such as cell-sorting and whole-genome sequencing.[68]

## Acknowledgements

This work was supported by National Institutes of Health (NIH) R35GM136223, R01AI141439 to J.X.C. Funding for the presented research was also provided via Young Independent Research Group Grant ZK-57 to F.C.P. M.W. was supported by the Wittgenstein award of the Austrian Science Fund FWF (Z-383B).

# Supplementary Information for

# Mid-Infrared Photothermal - Fluorescence in Situ Hybridization for Functional Analysis and Genetic Identification of Single Cells

Yeran Bai[1,3*], Zhongyue Guo[2,3*], Fátima C. Pereira[4], Michael Wagner[4,5†], Ji-Xin Cheng[1,2,3†]
[1]Department of Electrical and Computer Engineering, Boston University, Boston, MA 02215, USA

[2]Department of Biomedical Engineering, Boston University, Boston, MA 02215, USA

[3]Photonics Center, Boston University, Boston, MA 02215, USA
[3]Centre for Microbiology and Environmental Systems Science, Department of Microbiology and Ecosystem Science, University of Vienna, 1090 Vienna, Austria

[4]Department of Chemistry and Bioscience, Aalborg University, 9220 Aalborg, Denmark

* These authors contributed equally to this work.

†Corresponding authors. Emails: michael.wagner@univie.ac.at, jxcheng@bu.edu

This pdf file includes:
**Figure S1**. Data processing pipeline for high-throughput single-cell metabolic analysis from two IR wavenumber widefield MIP imaging.
**Figure S2**. Widefield MIP spectra of $^{12}$C- and $^{13}$C-glucose incubated *E. coli* cells.
**Figure S3**. Ultrafast imaging of 500 nm PMMA beads by the optimized widefield MIP setup.
**Figure S4**. Multispectral wide-field MIP imaging of standard samples and mock replacement ratio peaks.
**Figure S5**. Influence of FISH protocol on $^{13}$C-protein replacement ratio measurements of $^{12}$C-glucose incubated *E. coli* cells.
**Figure S6**. MIP-FISH imaging of an *E. coli* and *B. theta* mixture.
**Figure S7**. MIP imaging of a larger number of isotopically unlabelled human gut microbiome cells.
**Table S1**. FISH probe for single cell analysis.
**Table S2**. Reference coefficients used to calculate $^{13}$C-protein replacement ratio and statistics on reported experiments.
**Table S3**. The coefficients obtained for the same reference sample on different days.
**Table S4**. Mock ratio of standard samples.
**Supplementary Methods**. Fluorescence *in-situ* hybridization. Gut microbiome incubation.

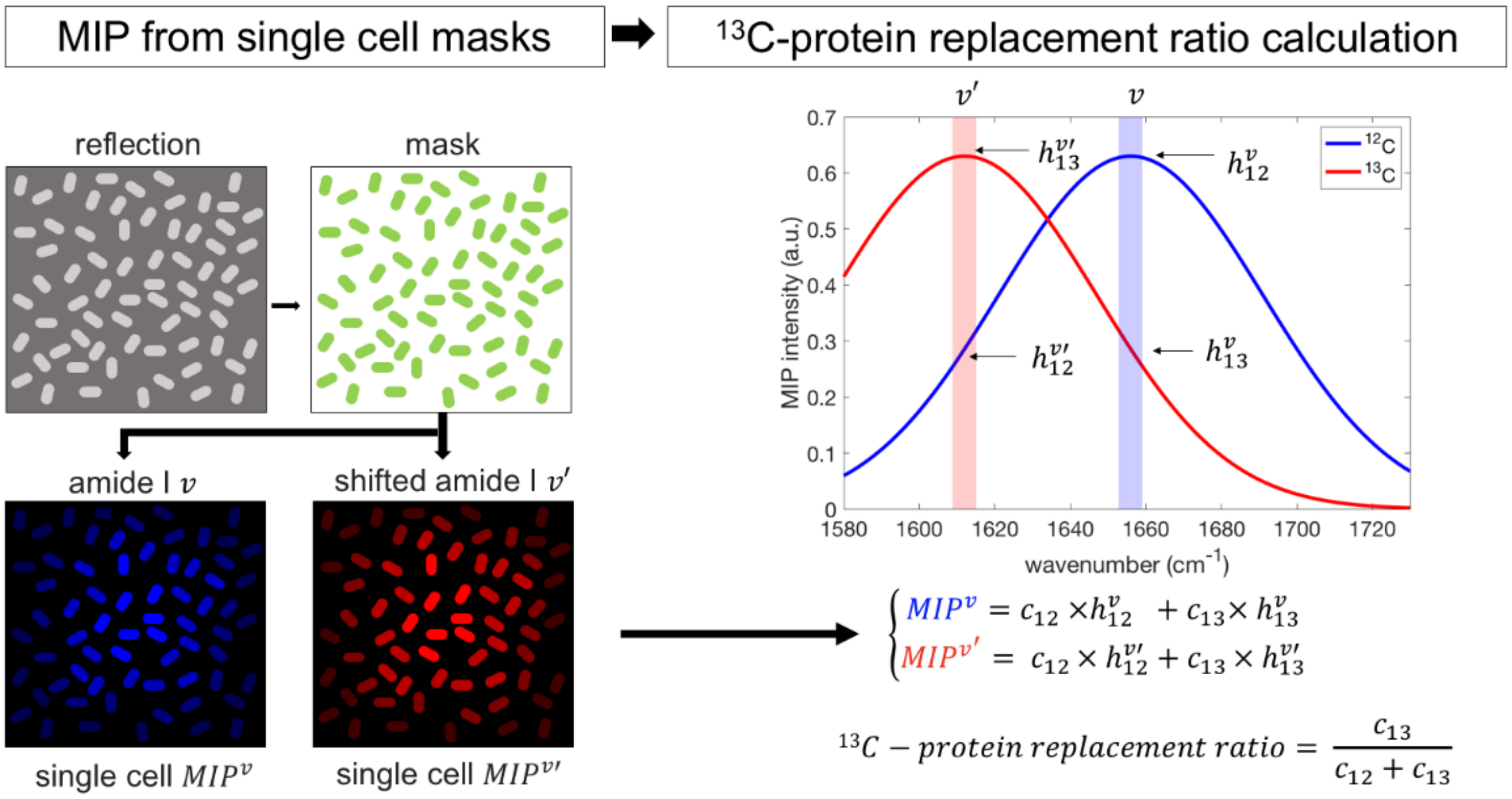

**Figure S1. Data processing pipeline for high-throughput single-cell metabolic analysis from two IR wavenumber widefield MIP imaging.** We imaged bacteria cells at two IR wavenumbers $\nu$ and $\nu'$ (centers around 1656 and 1612 $cm^{-1}$) for MIP imaging. Individual cells were selected based on the reflection images and a corresponding mask was created. The mask was applied to MIP images to measure MIP intensities for each cell. A simulated protein amide I region for $^{12}C$-glucose (blue curve) and $^{13}C$-glucose (red curve) incubated cells are shown on the right, with the $\nu$ and $\nu'$ indicated in blue and red shaded lines. The measured MIP intensities can be considered as a linear combination of $^{12}C$-protein and $^{13}C$-protein. We quantified the contribution from $^{12}C$-protein and $^{13}C$-protein ($c_{12}$ and $c_{13}$) to the measured MIP intensities at $\nu$ and $\nu'$ ($MIP^{\nu}$and $MIP^{\nu'}$) based on the coefficients ($h_{12}^{\nu}, h_{12}^{\nu'}, h_{13}^{\nu}, h_{13}^{\nu'}$ ) obtained from the reference sample. The $^{13}C$-protein replacement ratio is defined as $c_{13}/(c_{12} + c_{13})$. The coefficients used in the reported experiments are listed in Table S2.

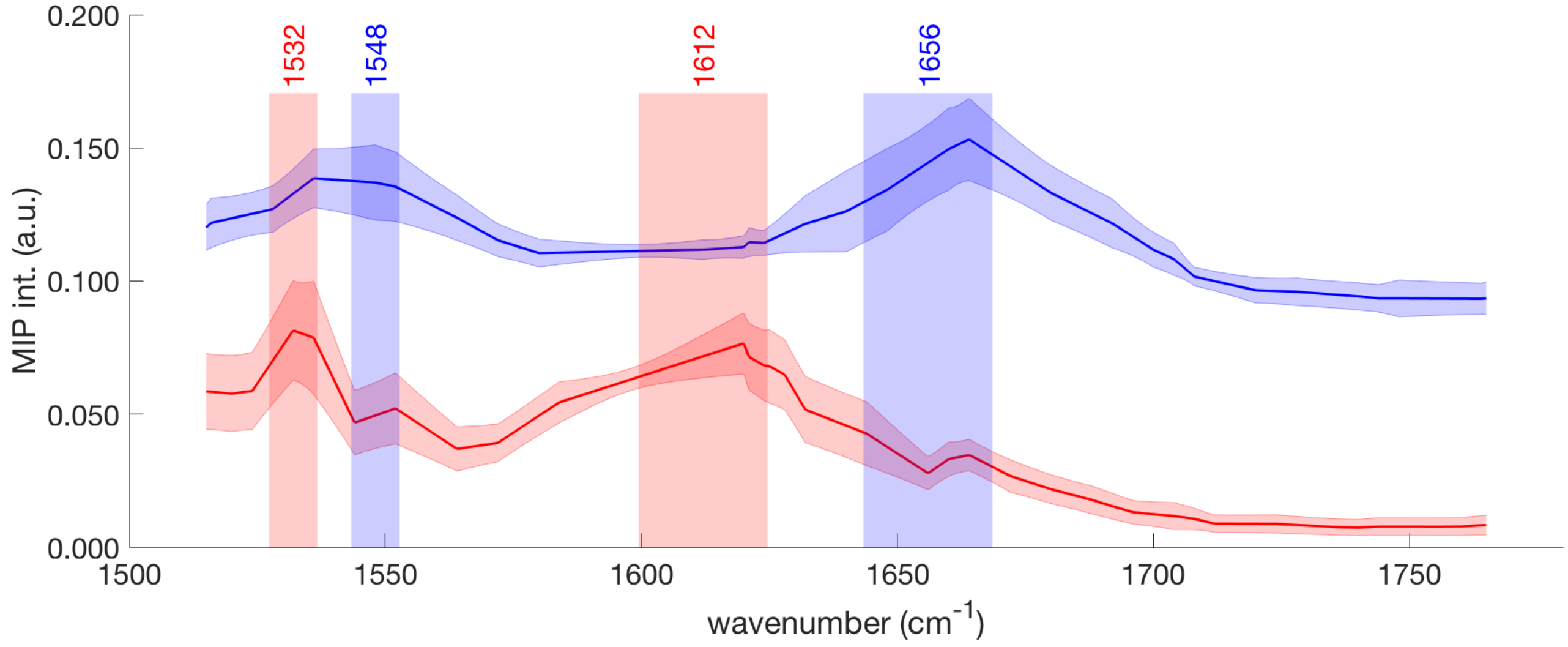

**Figure S2. Widefield MIP spectra of $^{12}$C- and $^{13}$C-glucose incubated *E. coli* cells.** The cells were incubated with $^{12}$C-glucose (blue) or $^{13}$C-glucose (red) for 24 hours. MIP images were acquired by tuning the IR laser in the 1512 to 1768 cm$^{-1}$ range with a step size around 10 cm$^{-1}$. Spectra were linear interpolated and offsetted to better visualize the trend. Standard deviation (shaded curve) of the spectra was determined from >20 single cells. The isotopic effect on protein spectra is clearly visualized by red-shifted peak (blue to red line vertical bars). For $^{13}$C-glucose incubated cells, a small residue peak can still be observed in the original peak positions for both amide I and amide II bands. However, the area under this amide I residue peak is only 4.4% of the area of this broad peak in unlabelled *E. coli* cells, indicating a near-full substitution of $^{13}$C in the total protein in *E.coli* cells incubated with $^{13}$C-glucose for 24 hours.

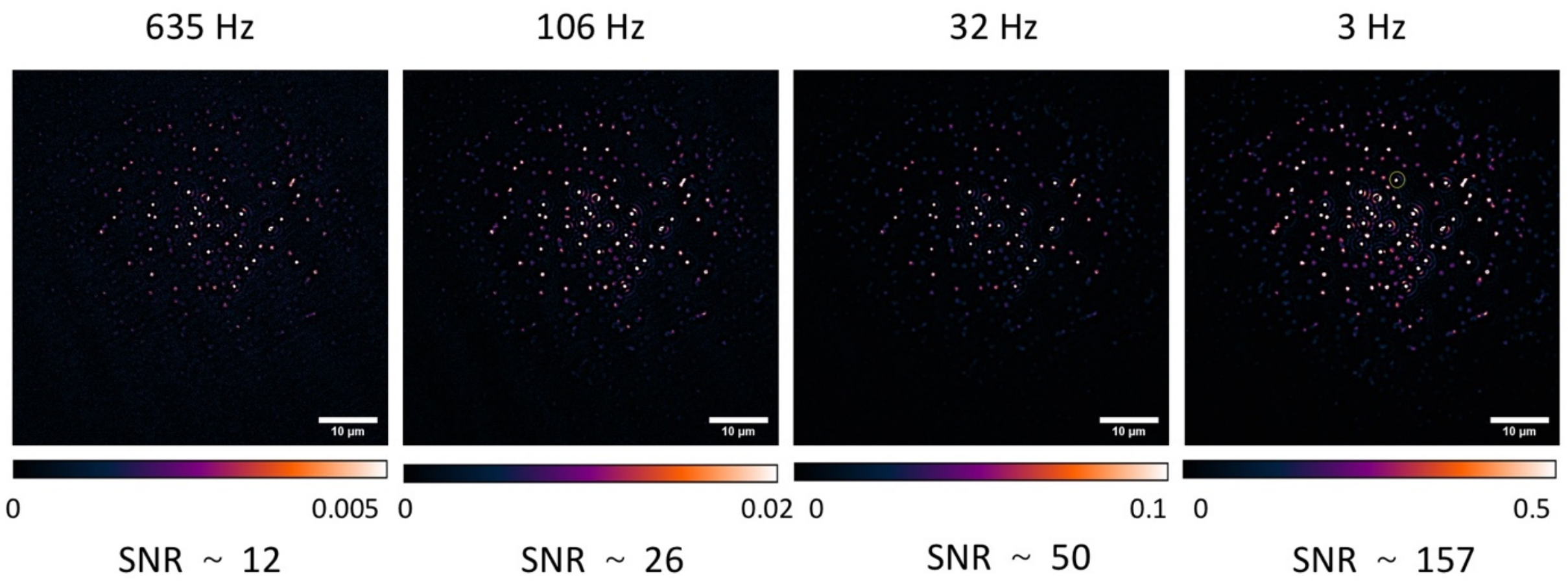

**Figure S3. Ultrafast imaging of 500 nm PMMA beads by the optimized widefield MIP setup.** PMMA beads were imaged at 1728 $cm^{-1}$ at different averaging frames. The signal to noise ratio (SNR) was calculated as the maximum MIP intensity of a single bead (circled in 3Hz image) divided by the standard deviation of a surrounding background area. A reasonable SNR ~12 was achieved for ultrafast imaging at 635 Hz, which corresponds to single 'IR on' – 'IR off' frame subtraction. Scale bar: 10 µm.

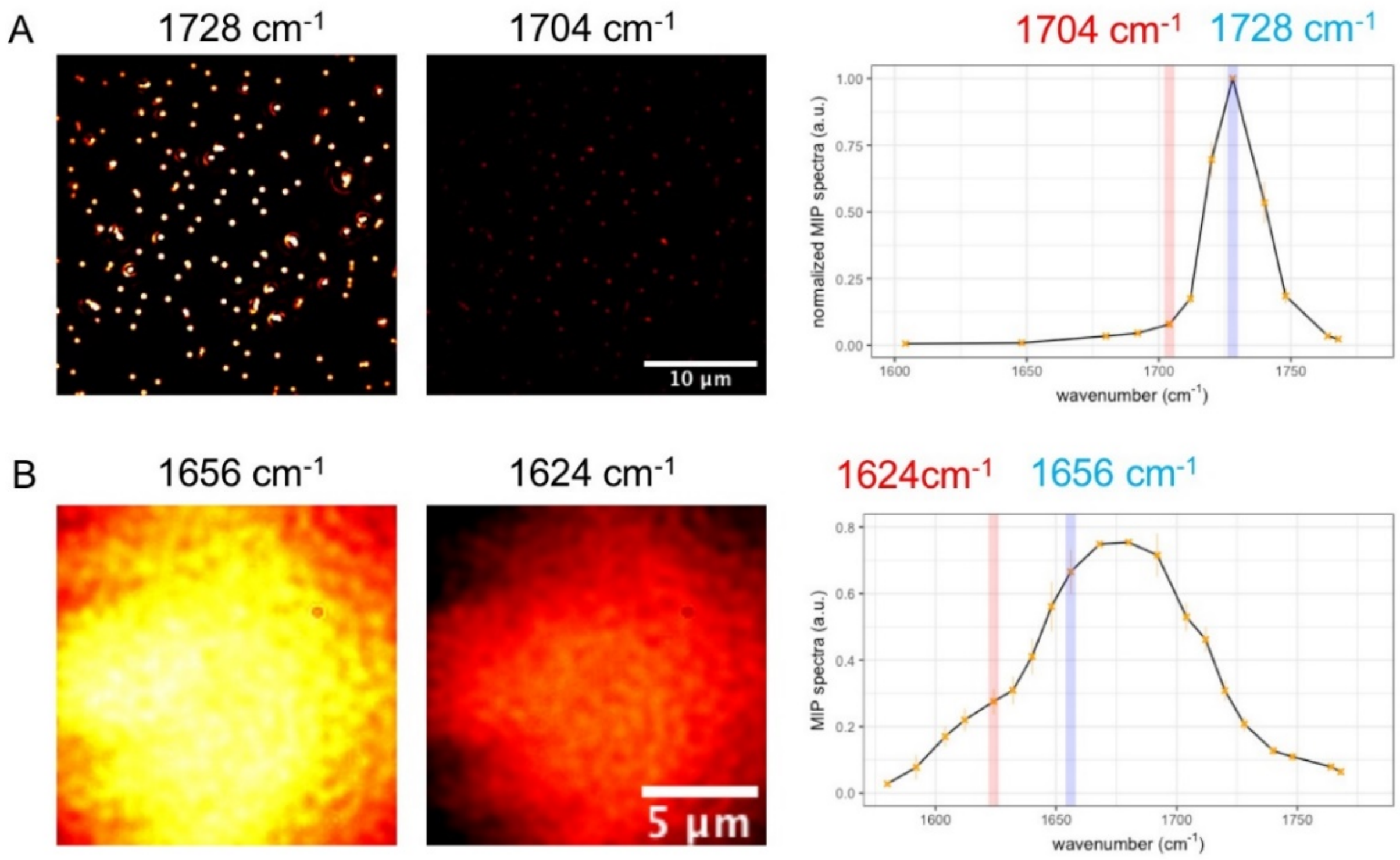

**Figure S4. Multispectral wide-field MIP imaging of standard samples and mock replacement ratio peaks.** For PMMA beads with 500 nm in diameter (A) and a bovine serum albumin (BSA) film (B), multispectral wide-field MIP images were acquired and the mock replacement ratio was calculated by applying an analogous analysis as for the $^{13}$C-protein replacement ratio (see **Figure S1**). For PMMA beads, 1728 cm$^{-1}$ and 1704 cm$^{-1}$ was used as the mock original and shifted amide I peak. For the BSA film, 1656 cm$^{-1}$ and 1624 cm$^{-1}$ was used as the mock original and shifted amide I peak. Wide-field MIP images at selected mock peaks are shown along with the widefield spectrum.

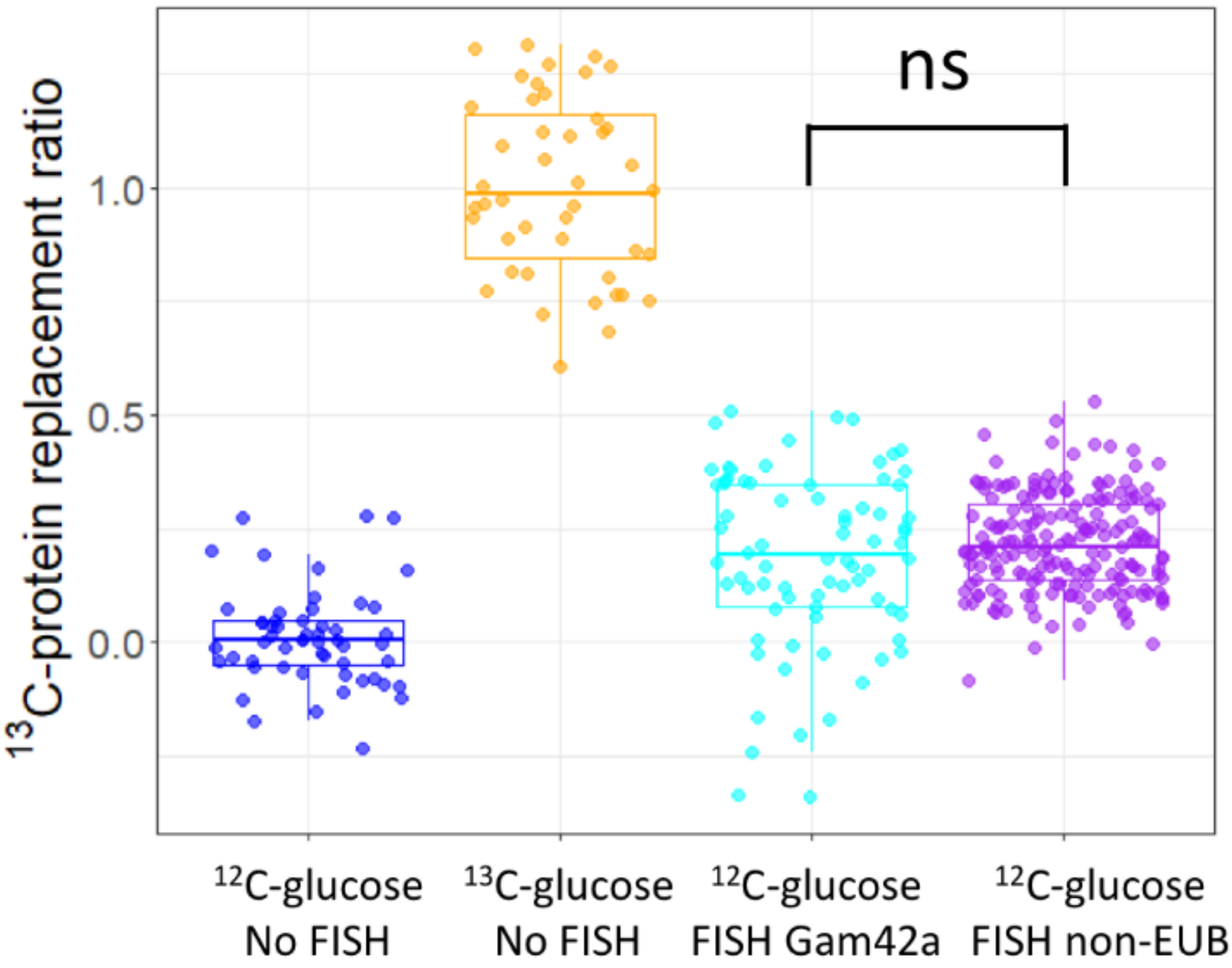

**Figure S5. Influence of FISH protocol on $^{13}C$-protein replacement ratio measurements of $^{12}C$-glucose incubated *E. coli* cells.** Measurements with $^{12}C$-glucose and $^{13}C$-glucose incubated and subsequently fixed *E.coli* cells, respectively that were not hybridized with fluorescently labelled oligonucleotide (No FISH) were used as references to infer the quantification coefficients. $^{12}C$-glucose incubated and subsequently fixed *E. coli* cells that were hybridized with the specific Gam42a FISH probe labelled with Cy5 or with a nonsense control FISH probe (non-EUB labelled with Cy5) showed a slight increase of the $^{13}C$-protein replacement ratio compared with the corresponding *E. coli* cells that were not hybridized. The $^{13}C$-protein replacement ratio inferred was not significantly different between Gam42a and non-EUB experiments (pairwise t-test, $p = 0.16$), suggesting that the hybridization or washing steps of the FISH protocol, but not the Cy5 fluorophore affect the quantification.

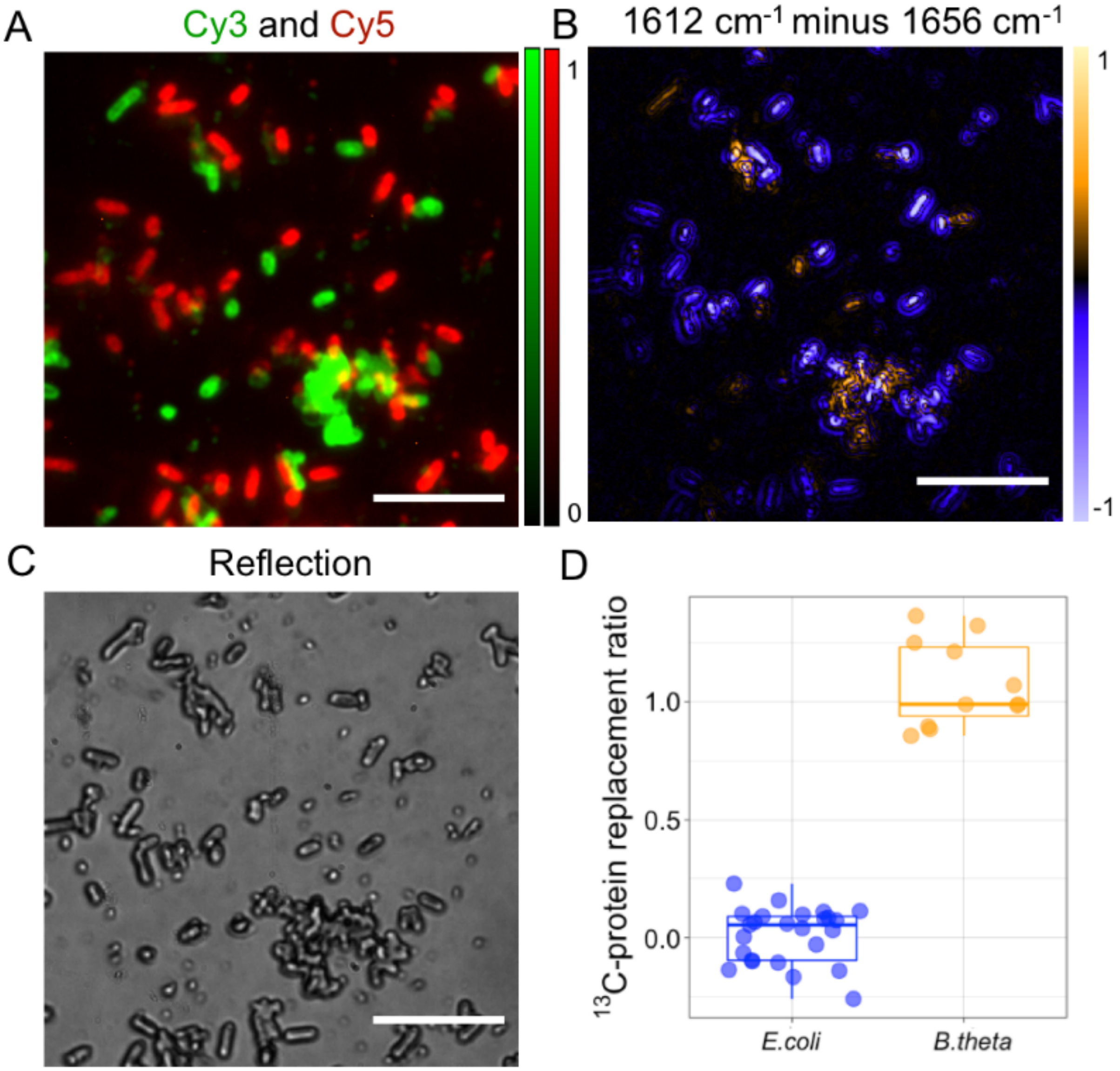

**Figure S6. MIP-FISH imaging of an *E. coli* and *B. theta* mixture.** In contrast to the experiment displayed in Figure 5, here *E. coli* cells were incubated with 0.4% (w/v) $^{12}$C-glucose and hybridized with Gam42a-Cy5 oligonucleotide probe, while *B. theta* cells grown in the presence of 0.5% (w/v) $^{13}$C-glucose were hybridized with a Bac303-Cy3 oligonucleotide probe. Subsequently, cells from both species were mixed and analyzed. (A) Fluorescence imaging for identification of *E. coli* (red) and *B. theta* (green). Scale bars 10 μm. (B) Subtraction of two MIP images (intensity at 1612 $cm^{-1}$ minus intensity at 1656 $cm^{-1}$) showed that a portion of the cells have incorporated $^{13}$C into the protein (in yellow) while other cells showed no $^{13}$C labelling (in purple). (C) Reflection image shows cell morphology of the bacterial mixture. (D) Quantification of $^{13}$C-protein replacement ratio. (Pairwise t-test: p=7.37e-11.)

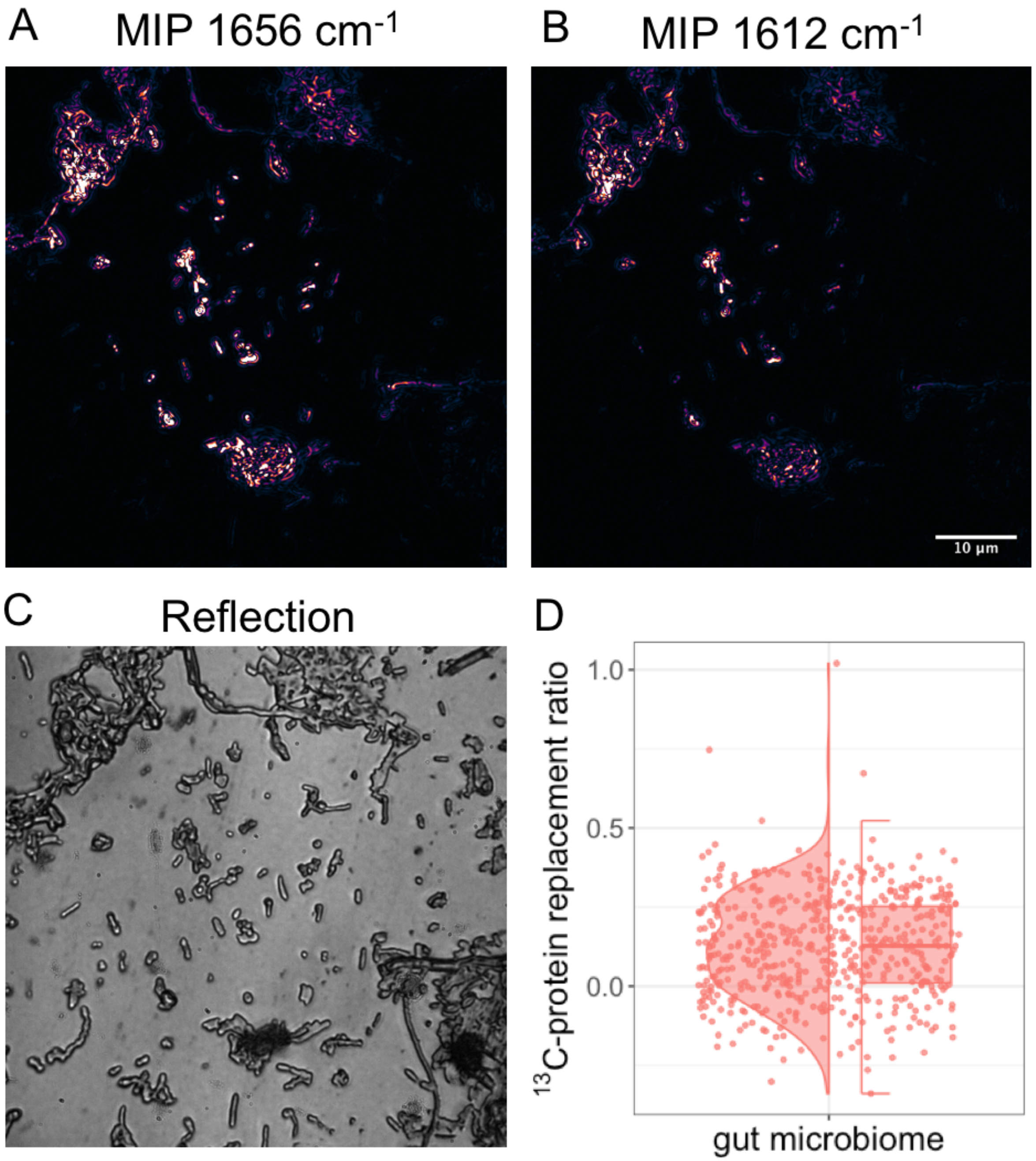

**Figure S7. MIP imaging of a larger number of isotopically unlabelled human gut microbiome cells.** Representative MIP and reflection images (A-C). (D) Distribution of 538 randomly selected cells showed a mean value of the $^{13}$C-protein replacement ratio of 0.127 and a standard deviation of 0.164.

**Table S1. FISH probe for single cell analysis.**

| Probe Name | Sequence (5' to 3') | Formamide (%) | Major target taxa* | Reference |
|---|---|---|---|---|
| **Gam42a** | GCCTTCCCACTT CGTTT | 35 | Class Gammaproteobacteria (28%) | (Manz, Amann, Ludwig, Wagner, & Schleifer, 1992) |
| **Bac303** | CCAATGTGGGG GACCTT | 10 | Family Prevotellaceae (73%)<br>Family Barnesiellaceae (57%)<br>Family Tannerellaceae (50%) | (Manz, Amann, Ludwig, Vancanneyt, & Schleifer, 1996) |
| **NON-EUB** | ACTCCTACGGG AGGCAGC | - | negative control probe | (Wallner, Amann, & Beisker, 1993) |

* According to Silva database v138.1 (https://www.arb-silva.de/). The probe coverage for each taxon is shown in parenthesis.

**Table S2. Reference coefficients used to calculate $^{13}$C-protein replacement ratio and statistics on reported experiments.**

| sample | Reference sample | Reference coefficients | Mean replacement ratio | standard deviation | number of ROIs |
|---|---|---|---|---|---|
| **Figure 3. High detection sensitivity for $^{13}$C-incorporation.** | | | | | |
| 0% $^{13}$C-glucose | *E. coli* 0% $^{13}$C-glucose *E. coli* 100% $^{13}$C-glucose | $h_{12}^{v}$ = 0.688, $h_{12}^{v'}$ = 0.311, $h_{13}^{v}$= 0.289, $h_{13}^{v'}$ = 0.711 | 0.01100 | 0.07815 | 332 |
| 5% $^{13}$C-glucose | | | 0.06691 | 0.07627 | 454 |
| 10% $^{13}$C-glucose | | | 0.14437 | 0.06573 | 449 |
| 30% $^{13}$C-glucose | | | 0.35804 | 0.06516 | 155 |
| 50% $^{13}$C-glucose | | | 0.51838 | 0.12283 | 274 |
| 70% $^{13}$C-glucose | | | 0.69620 | 0.15375 | 243 |
| 90% $^{13}$C-glucose | | | 0.88839 | 0.11189 | 231 |
| 100% $^{13}$C-glucose | | | 0.99900 | 0.09107 | 212 |
| **Figure 4. FISH is compatible with MIP metabolic imaging.** | | | | | |
| $^{12}$C-glucose, fixed cells, no FISH | *E. coli* $^{12}$C-glucose no FISH *E. coli* $^{13}$C-glucose no FISH | $h_{12}^{v}$ = 0.626, $h_{12}^{v'}$ = 0.374, $h_{13}^{v}$= 0.333, $h_{13}^{v'}$ = 0.667 | 0.01099 | 0.09424 | 115 |
| $^{13}$C-glucose, fixed cells, no FISH | | | 0.99900 | 0.12762 | 286 |
| $^{13}$C-glucose, Cy5-labelled cells | | | 0.92411 | 0.14720 | 553 |
| **Figure 5. MIP-FISH imaging of bacterial mixtures.** | | | | | |
| *E. coli* ($^{13}$C, Cy5) | *E. coli* ($^{12}$C, Cy5) *E. coli* ($^{13}$C, Cy5) | $h_{12}^{v}$ = 0.573, $h_{12}^{v'}$ = 0.427, $h_{13}^{v}$= 0.298, $h_{13}^{v'}$ = 0.702 | 0.99764 | 0.20615 | 71 |
| *B. theta* ($^{12}$C,Cy3) | *B. theta* ($^{12}$C,Cy3) *B. theta* ($^{13}$C,Cy3) | $h^{v}_{12}$ = 0.618, $h^{v'}_{12}$ = 0.382, $h^{v}_{13}$= 0.444, $h^{v'}_{13}$ = 0.556 | -0.00303 | 0.32610 | 53 |
| **Figure 6. MIP-FISH imaging of a gut microbiome sample with spiked $^{13}$C-labelled *E. coli* cells.** | | | | | |
| *E. coli* ($^{13}$C, Cy5) | *E. coli* 0% $^{13}$C-glucose *E. coli* 100% $^{13}$C-glucose | $h^{v}_{12}$ = 0.674, $h^{v'}_{12}$ = 0.326, $h^{v}_{13}$= 0.276, $h^{v'}_{13}$ = 0.723 | 0.94167 | 0.07274 | 42 |
| Gut microbiome | | | 0.12506 | 0.16209 | 593 |
| **Figure S5. Influence of FISH protocol on $^{13}$C-protein replacement ratio measurements of $^{12}$C-glucose incubated *E. coli* cells.** | | | | | |
| *E. coli* $^{12}$C no FISH | *E. coli* $^{12}$C no FISH *E. coli* $^{13}$C no FISH | $h^{v}_{12}$ = 0.674, $h^{v'}_{12}$ = 0.326, $h^{v}_{13}$= 0.392, $h^{v'}_{13}$ = 0.608 | 0.01100 | 0.10652 | 55 |
| *E. coli* $^{13}$C no FISH | | | 0.99900 | 0.19486 | 44 |
| *E. coli* $^{12}$C FISH Gam42a | | | 0.18481 | 0.19354 | 76 |
| *E. coli* $^{12}$C FISH non-EUB | | | 0.21810 | 0.10642 | 191 |
| **Figure S6. MIP-FISH imaging of an *E. coli* and *B. theta* mixture.** | | | | | |
| *E. coli* ($^{12}$C, Cy5) | *E. coli* ($^{12}$C, Cy5) *E. coli* ($^{13}$C, Cy5) | $h^{v}_{12}$ = 0.573, $h^{v'}_{12}$ = 0.427, $h^{v}_{13}$= 0.298, $h^{v'}_{13}$ = 0.702 | 0.01268 | 0.11680 | 25 |
| *B. theta* ($^{13}$C,Cy3) | *B. theta* ($^{12}$C,Cy3) *B. theta* ($^{13}$C,Cy3) | $h^{v}_{12}$ = 0.618, $h^{v'}_{12}$ = 0.382, $h^{v}_{13}$= 0.444, $h^{v'}_{13}$ = 0.556 | 1.07379 | 0.18268 | 11 |
| **Figure S7. MIP imaging of a larger number of isotopically unlabelled human gut microbiome cells.** | | | | | |
| Gut microbiome | *E. coli* 0% $^{13}$C-glucose *E. coli* | $h^{v}_{12}$ = 0.674, $h^{v'}_{12}$ = 0.326, | 0.12696 | 0.16374 | 538 |
| | 100% $^{13}C$-glucose | $h_{13}^{v}$= 0.276, $h_{13}^{v'}$ = 0.723 | | | |

**Table S3. The coefficients obtained for the same reference sample on different days.** The reference sample here was the *E. coli* cells incubated with 0.2% (w/v) $^{12}C$-glucose or $^{13}C$-glucose for 24 hours and subsequently fixed but not hybridized, reported in Figure 3.

| | **Day 1** | **Day 2** | **Day 3** |
|---|---|---|---|
| $h_{12}^{v}$ | 0.688 | 0.691 | 0.674 |
| $h_{12}^{v'}$ | 0.311 | 0.309 | 0.326 |
| $h_{13}^{v}$ | 0.289 | 0.298 | 0.276 |
| $h_{13}^{v'}$ | 0.711 | 0.702 | 0.723 |

**Table S4. Mock ratio of standard samples.** The acquisition time was 0.31 second per IR wavenumber.

| **Sample** | **Normal peak (cm-1)** | **Shifted peak (cm-1)** | **Mean replacement ratio** | **Standard deviation** | **Number of ROIs** |
|---|---|---|---|---|---|
| PMMA beads (500nm in diameter) | 1728 | 1704 | 0.01299 (mock) | 0.00895 | 40 |
| BSA film | 1656 | 1624 | 0.13313 (mock) | 0.01553 | 72 |

## Supplementary Methods

**Fluorescence *in-situ* hybridization.** Fixed cells (100 µL) were pelleted at 14,000 × *g* for 10 min, resuspended in 100 µL 96% analytical grade ethanol, and incubated for 1 min at room temperature for dehydration. Subsequently, the samples were centrifuged at 14,000 × *g* for 5 min, the ethanol was removed, and the cell pellet was air-dried. Cells were hybridized in solution (100 µL) for 3 h at 46 °C. The hybridization buffer consisted of 900 mM NaCl, 20 mM Tris-(hydroxymethyl)-amino methane HCl, 1 mM ethylenediamine tetraacetic acid, and 0.01% sodium dodecylsulphate and contained 100 ng of the respective fluorescently labelled oligonucleotide as well as the required formamide concentration to obtain stringent conditions (**Table S1**). After hybridization, samples were immediately transferred into a centrifuge with a rotor preheated at 46 °C and centrifuged at 14,000 × *g* for 15 min at the maximum allowed temperature (40 °C) to minimize unspecific probe binding. Samples were washed in a buffer of appropriate stringency (Daims, Stoecker & Wagner, 2006) for 15 min at 48 °C, and cells were centrifuged for 15 min at 14,000 × *g* at the maximum allowed temperature (40 °C) in a centrifuge with a rotor preheated at 46 °C. Cells were finally washed with 500 µL of ice-cold phosphate buffered saline, resuspended in 20 µL of phosphate buffered saline and stored at 4 °C until further use.

**Gut microbiome incubation.** Human fecal samples were collected from six healthy adult individuals (two male and four females between the ages of 26 to 39) who had not received antibiotics in the prior 3 months. Study participants provided informed consent and self-sampled using an adhesive paper-based feces catcher (FecesCatcher, Tag Hemi, Zeijen, NL) and a sterile polypropylene tube with the attached sampling spoon (Sarstedt, Nümbrecht, DE). The study protocol was approved by the University of Vienna Ethics Committee (reference No.00161). All meta(data) is 100% anonymized and compliant with the University's regulations. Samples were transferred into an anaerobic tent (Coy Laboratory Products, USA) within 30 min after sampling. Samples were suspended in M9 medium (prepared without glucose) to achieve a concentration of 0.1 g/ml, left to settle for 10 minutes, and the supernatants were combined. The combined fecal slurry was further diluted 10 times in this medium. The homogenate was left to settle for 10 minutes, and the supernatant was then distributed into glass vials. Each vial was supplemented with D-glucose (5 mg/ml; unlabelled D-Glucose, 99.5%, Sigma-Aldrich). After incubation for 6 h at 37 °C under anaerobic conditions (5% $H_2$, 10% $CO_2$, 85% $N_2$), sample aliquots were collected by centrifugation. Aliquots were fixed in 4% formaldehyde for 2 h at 4°C. Samples were finally washed two times with 1 ml of PBS and stored in PBS at 4 °C until further use.